\documentclass[openacc]{rsproca_new}

\usepackage{amsmath}
	\DeclareMathOperator*{\argmin}{arg\,min}
\usepackage{amssymb}
\usepackage{graphics}
\usepackage{listings}
\usepackage{algorithm}
\usepackage{algpseudocode}
\usepackage{xcolor}

\begin{document}

\title{Adaptive Bayesian Inference of Markov Transition Rates}

\author{
Nicholas W. Barendregt$^{1}$, Emily G. Webb$^{2}$ and Zachary P. Kilpatrick$^{1}$}

\address{$^{1}$Department of Applied Mathematics, University of Colorado Boulder, 1111 Engineering Center, ECOT 225, 526 UCB, Boulder, Colorado, 80309, USA\\
$^{2}$Applied Physics Laboratory, Johns Hopkins University, 11100 Johns Hopkins Road, Laurel, Maryland, 20723, USA}

\subject{applied mathematics, statistics}


\corres{Nicholas W. Barendregt\\
\email{nicholas.barendregt@colorado.edu}}

\begin{abstract}
Optimal designs minimize the number of experimental runs (samples) needed to accurately estimate model parameters, resulting in algorithms that, for instance, efficiently minimize parameter estimate variance. Governed by knowledge of past observations, adaptive approaches adjust sampling constraints online as model parameter estimates are refined, continually maximizing expected information gained or variance reduced. We apply adaptive Bayesian inference to estimate transition rates of Markov chains, a common class of models for stochastic processes in nature. Unlike most previous studies, our sequential Bayesian optimal design is updated with each observation, and can be simply extended beyond two-state models to birth-death processes and multistate models. By iteratively finding the best time to obtain each sample, our adaptive algorithm maximally reduces variance, resulting in lower overall error in ground truth parameter estimates across a wide range of Markov chain parameterizations and conformations.
\end{abstract}

\maketitle

\section{Introduction}
Using experimental data to infer parameters is essential for accurate quantitative models of natural phenomena.
Inherent stochasticity in most physical systems compounds this difficulty, clouding the link between data and ground truth in ways experimentalists cannot control. Not only does noise cause uncertainty in model parameter estimates, but it can slow the process of model refinement.
As a result, researchers historically utilized statistical methods to design experiments that maximize the information obtained from each experimental measurement~\cite{lindley1956,nishii1993,johnson2011}.
In particular, Bayesian experimental design, applied to a system with unknown parameters $\mathbf{x}$, starts with a prior $p(\mathbf{x})$, constructs the associated posterior $p(\mathbf{x}\vert \Theta, \Xi)$ based on data $\Theta$ obtained from an experimental design $\Xi$, and finds the design that optimizes a specified objective, such as minimizing a utility function (e.g.,~variance) that incorporates sampling-associated costs (e.g.,~time or resources needed to take measurements)~\cite{chaloner1995,ryan2016}.
These methods have seen wide application in economics \cite{bandi2008}, queueing theory \cite{ehrenfeld1962}, physics \cite{huan2013,dushenko2020}, and cognitive neuroscience \cite{myung2013,cavagnaro2010}. 
In particular, Bayesian experimental design for inferring transition rates in discrete-state Markov models has seen great success when applied to simple epidemiological~\cite{cook2008,ross2009,ferguson2014} and ecological~\cite{becker1983,pagendam2010} models. Recent efforts have shown that adaptive designs can speed up the timescale of clinical drug or intervention trials~\cite{bhatt2016}, providing an automated, model-based prescription that governs future sampling.

One of the outputs of Bayesian experimental designs, for systems producing time series data, is a sampling schedule, a set of times to measure the state of a system, chosen to optimize an objective function (e.g., minimizing sample number for a fixed estimate tolerance, maximizing sample information).
When schedules are planned in advance of experiments, they may require sampling continuously in time or periodically with a fixed sampling frequency~\cite{mehtala2015}, which may be infeasible or inefficient given high sampling costs.
For example, an ecologist studying the dynamics of several interacting species may be restricted by seasonal patterns of animal activity, the expense or time cost of field work, or a finite project timeline, such that they are unable to implement a fixed schedule of sampling population sizes.
In these situations, Bayesian experimental design can be extended to incorporate sequential analysis~\cite{chernoff1959}, yielding iterative and adaptive sampling schedules based on prior observations.
While approximate versions of these Bayesian adaptive designs have been applied to simple models~\cite{michel2020}, precise sequential formulations applied to more complicated systems are often limited by intractable likelihood functions. 
This difficulty has spawned advanced algorithms involving stochastic optimization~\cite{pagendam2009,huan2014}, Markov Chain Monte Carlo (MCMC)~\cite{drovandi2013,drovandi2014,ryan2016} and machine learning methods~\cite{rainforth2017,foster2021}.

In this work, we develop an adaptive sequential Bayesian inference algorithm that successively optimizes each process sample time to minimize the variance of transition rate parameter estimates for discrete-state Markov processes with arbitrary numbers of states and transition rates. An early version of this work focused on simple two-state processes with a single transition rate~\cite{webb2021}.
Starting with two-state Markov chains, we illustrate how sequentially chosen sampling times are selected to minimize expected parametric posterior variance after each observation.
We compare the speed and accuracy of this adaptive algorithm to that of a fixed-period sampling algorithm across transition rate parameter space. Considering more complex Markov chains, our algorithm can be extended by minimizing the expected determinant of the covariance matrix associated with the transition rate matrix.
We apply this algorithm to three specific Markov chain models: a ring of states, modeling the diffusive degradation of memory for a single circular parameter~\cite{wimmer2014bump,panichello2019}, a birth-death process describing population dynamics of epidemics and ecological groups~\cite{cook2008}, and a general Markov chain inference problem with a binarized prior for each transition rate.
Taken collectively, these results demonstrate a simple yet powerful approach to efficiently inferring the dynamics of Markov models.

\section{Methods: Adaptive Bayesian Design for Markov Chains}
	\label{sec:Simple_Networks}
	Our Bayesian inferential approach to determining transition rates of a Markov chain from time series observations relies on obtaining accurate representations of the parametric likelihood functions from state observations. 
	This is plausible in the case of simple Markov chains for which likelihoods can be determined analytically. 
	Of course, for continuous-time Markov chains, it is always possible to write down likelihood formulas, but computation becomes infeasible for sufficiently large chains. 
	In this section, we derive our algorithm for adaptive Bayesian inference through a series of examples with increasing complexity.
	These examples culminate with our derivation of the adaptive algorithm for general time-homogeneous, continuous-time Markov chains.
	\subsection{Inferring Single Transition Rates}
		We start by considering a continuous-time Markov process with two states $X(t)\in\{0,1\}$ and a single transition rate $h_0>0$, so that 
		\begin{equation}
			\begin{gathered}
				\Pr\left(X(t+\delta t)=1\vert X(t)=0\right) =h_0\delta t+o\left(\delta t\right), \quad \delta t\downarrow0,  \label{unidir} \\
				\Pr(X(t+\delta t)=0\vert X(t)=1)=0, \quad \forall \delta t\ge0.
			\end{gathered}
		\end{equation}
		For this Markov process with a single transition link, we will enforce the initial condition $X(0)=0$ to guarantee the possibility of observing a transition from $X=0$ to $X=1$.
		Our goal is to design an algorithm for sampling from this process efficiently to optimally decrease the estimate variance of the transition rate parameter $h_0$ with each state sample (after reinitialization).
		Using Bayesian inference, we assume a prior distribution $p_0(h_0)$ and obtain measurements of the process $\xi_i=\left(t_i,X(t_i)\right)=\left(t_i,X_i\right)$, where $i\in\{0,1,2,\dots\}$ represents the sample number index and $X$ evolves according to Eq.~\eqref{unidir} and always taking $X(0) = 0$. 
		These samples allow us to construct the posterior $p_n(h_0)=  p(h_0\vert\xi_{1:n})$.
		Sampling times $t_i$ are chosen to minimize the expected posterior variance after each observation.
		The algorithm samples observations until a predetermined threshold variance $\theta$ is reached. At this point, the transition rate estimator $\hat{h}_0$ is the maximum a posteriori estimator.
	
		To illustrate the process of selecting each sample time $t_n$, suppose we have have previously observed $n-1$ measurements, $\xi_{1:n-1} = \left\{ (t_1,X_1), ... (t_{n-1}, X_{n-1}) \right\}$. 
		Given a particular planned subsequent observation time $t_n$, the expected posterior variance on the next ($n$th) timestep  $ \overline{\rm Var}_n ( h_0 \vert \xi_{1:n-1}, t_n) $ is given by marginalizing over the possible future measurements $\xi_n$ (i.e., possible state observations $X(t_n)$) assuming the history of observations $\xi_{1:n-1}$ and a sample time $t_n$:
		\begin{align*}
			\overline{\rm Var}_n(h_0 \vert \xi_{1:n-1}, t_n)=&  {\rm Var}\left(h_0\vert X_n=0, t_n, \xi_{1:n-1} \right)\Pr\left(X_n =0 \vert t_n, \xi_{1:n-1} \right) \\ & +{\rm Var}\left(h_0\vert X_n=1, t_n, \xi_{1:n-1} \right)\Pr\left(X_n=1 \vert t_n , \xi_{1:n-1} \right).
		\end{align*}
		Note, that since we always reset $X(0) = 0$ preceding each sample, the relevant conditional observation probabilities are
		\begin{equation}
			p(\xi_n \vert h_0 )=\begin{cases}
				1-e^{-h_0t_n}, & X(t_n) = 1 \\
				e^{-h_0t_n}, & X(t_n) = 0
			\end{cases},
		\end{equation}
		and 
		\begin{equation*}
			p(\xi_n \vert \xi_{1:n-1}) = \int_0^{\infty} p(\xi_n \vert h_0) p_{n-1}(h_0) d h_0.
		\end{equation*}
		Thus, we obtain the expected variance formula
		\begin{multline}
			\overline{\text{Var}}_n(h_0 \vert \xi_{1:n-1}, t_n)=\int_0^\infty p_{n-1}(h_0) \Bigg[e^{-h_0t_n}\left\{h_0-\frac{\int_0^\infty h_0e^{-h_0t_n}p_{n-1}(h_0)\,dh_0}{\int_0^\infty e^{-h_0t_n}p_{n-1}(h_0)\,dh_0}\right\}^2\\
			+\left(1-e^{-h_0t_n}\right)\left\{h_0-\frac{\int_0^\infty h_0\left(1-e^{-h_0t_n}\right)p_{n-1}(h_0)\,dh_0}{\int_0^\infty\left(1-e^{-h_0t_n}\right)p_{n-1}(h_0)\,dh_0}\right\}^2\Bigg]\,dh_0.
			\label{eq:two-state_unidirectional_variance}
		\end{multline}
		The sample time $t_n$ that minimizes Eq.~\eqref{eq:two-state_unidirectional_variance} depends on the posterior $p_{n-1}$, computed from the sequence of previous observations of the stochastic process.
		Because each sample time depends on the previous posterior distribution, this algorithm performs sequential and \emph{adaptive Bayesian inference}.
		The expected variance formula in Eq.~\eqref{eq:two-state_unidirectional_variance} can be defined iteratively. Moreover, once $t_n$ is chosen by minimizing Eq.~\eqref{eq:two-state_unidirectional_variance} and an observation is made for $X_n$, we can calculate the true variance after the $n$th observation as
		\begin{align}
			\text{Var}_n(h_0) = \int_0^{\infty} p_n(h_0)\left\{h_0- \int_0^\infty h_0 p_{n}(h_0)\,dh_0 \right\}^2 d h_0,
		\end{align}
		where 
		\begin{equation*}
			p_n(h_0) = p_{n-1}(h_0) \cdot \begin{cases} 1 - e^{-h_0 t_n}, & X(t_n) = 1 \\ e^{-h_0t_n}, & X(t_n) = 0. \end{cases}.
		\end{equation*}
		This leads us to propose Algorithm~\ref{alg:two-state_unidirectional}.
		Unless otherwise noted, we will take the variance threshold, which terminates the accumulation of observations, to be $\theta=0.1$ throughout this work.
		\begin{algorithm}[t]
			\caption{Single-transition adaptive Bayesian inference}
			\begin{algorithmic}
				\Require $n=0$, $\theta > 0$, $p_0(h_0)$ \Comment{$p_0$: prior with support $[0,\infty)$.}
				\While{$\text{Var}_n(h_0) > \theta$}
				\State $n \gets n+1$
				\State $t_n\gets\argmin_{t\ge0}\overline{\text{Var}}_{n}(h_0;t)$ \Comment{Calculate $\overline{\text{Var}}_{n}$ using Eq.~\eqref{eq:two-state_unidirectional_variance}.}
				\State Draw $\xi_n=(t_n , X_n)$
				\State $p_n(h_0)\gets \frac{p(\xi_n\vert h_0)p_{n-1}(h_0)}{\int_0^\infty p(\xi_n\vert h_0)p_{n-1}(h_0)\,dh_0}$
				\EndWhile
			\end{algorithmic}
			\label{alg:two-state_unidirectional}
		\end{algorithm}

	\subsection{Multi-dimensional Inference on Simple Chains}
		\label{sec:Simple_Networks_Multiple_Rates}
		Our adaptive inference algorithm easily extends to chains with multiple transition rates, as we simply need to compute the state probability distribution for the Markov chain and include that in our Bayesian update. To illustrate, consider the same two-state Markov process, but with transitions occurring bidirectionally with transition rates $h_0$ ($0 \mapsto 1$) and $h_1$ ($1 \mapsto 0$).
		Expanding the inference problem beyond a single dimension requires defining a new objective function to minimize, which will now involve multiple transition rate parameters: variability in the estimate is now defined by the posterior covariance matrix $\Sigma$ rather than the variance.
		There are several ways to ``minimize'' a covariance matrix~\cite{nishii1993,jones2020}. 
		Here, we take the approach of minimizing the determinant of the expected covariance, known as a ``D-optimal'' method in optimal experimental design. 
		Such an approach is also equivalent to maximizing the product of the eigenvalues of the Fisher information matrix~\cite{deAguiar1995}. 
		Maximizing information gain here is preferable to reducing averaging variance (as in A-optimal designs), since there could be strong asymmetry in the transition rate parameters.
	
		The process is always guaranteed to eventually switch from one state to another as long as both rates are nonzero, and the transition rate parameters can both be inferred to arbitrarily small variances given enough observations. 
		Thus, we avoid the need to reset the chain's state after each observation.
		For simplicity, we assume $X_0 = X(t_0=0) = 0$ to begin, but it is not difficult to extend the algorithm to the case where $X_0$ is chosen randomly, and we subsequently allow the variable to evolve according to a Markov chain whose transition rates are chosen from the prior, $(h_0, h_1) \sim p_0(h_0,h_1)$. Thereafter, $X_n = X(t_n)$ is drawn and compared with $X_{n-1} = X(t_{n-1})$ to update the posterior over the transition rates $(h_0, h_1)$.
	
		As with the adaptive inference procedure for a single transition rate, we determine the next sample time $t_n$ after the current time $t_{n-1}$ by minimizing the determinant of the expected covariance matrix. For an arbitrary subsequent sampling time $t_n$, the expected covariance $\left[\Sigma_n\right]_{ij} \equiv \overline{\text{Cov}}_n(h_i,h_j)$ is computed by marginalizing over the possible observations $X_n$ and conditioning on the past observations $\xi_{1:n-1}$:
		\begin{align*}
			\overline{\text{Cov}}_n(h_i,h_j)=&\text{Cov}_n(h_i,h_j\vert X_n=0, t_n, \xi_{1:n-1})p(X_n=0 \vert t_n, \xi_{1:n-1}) \\
			& +\text{Cov}_n(h_i,h_j\vert X_n=1, t_n, \xi_{1:n-1}) p(X_n=1\vert t_n, \xi_{1:n-1}).
		\end{align*}
		Now, marginalizing over transition probabilities from the previous state $X_{n-1}$, which we can define for all possible cases
		\begin{equation}
			p(X_n=j\vert X_{n-1}=i)=\begin{cases}
				\frac{h_1}{h_0+h_1}+\frac{h_0}{h_0+h_1}e^{-(h_0+h_1)(t_n-t_{n-1})}, & i=j=0\\
				\frac{h_0}{h_0+h_1}-\frac{h_0}{h_0+h_1}e^{-(h_0+h_1)(t_n-t_{n-1})}, & i=0,\ j=1\\
				\frac{h_1}{h_0+h_1}-\frac{h_1}{h_0+h_1}e^{-(h_0+h_1)(t_n-t_{n-1})}, & i=1,\ j=0\\
				\frac{h_0}{h_0+h_1}+\frac{h_1}{h_0+h_1}e^{-(h_0+h_1)(t_n-t_{n-1})}, & i=j=1
			\end{cases},
			\label{eq:two-state_bidirectional_transition_probs}
		\end{equation}
		yields the expected future covariance
		\begin{equation}
			\begin{aligned}
				{}&\overline{\text{Cov}}_n(h_i,h_j;t_n)=\\
				{}& \sum_{k=0}^1 \iint_{\mathbb{R}_{\geq 0}^2} \Bigg[\left\{h_i-\frac{\int_0^\infty h_i\left( \int_0^{\infty} p(X_n=k\vert X_{n-1})p_{n-1}(h_0,h_1)^2\,dh_j\right)\,dh_i}{\iint_{\mathbb{R}_{\geq 0}^2} p(X_n=k\vert X_{n-1})p_{n-1}(h_0,h_1)^2\,dh_0\,dh_1}\right\}\\
				\times{}&\left\{h_j-\frac{\int_0^\infty h_j\left(\int_0^\infty p(X_n=k\vert X_{n-1})p_{n-1}(h_0,h_1)^2\,dh_i\right)\,dh_j}{\iint_{\mathbb{R}_{\geq 0}^2} p(X_n=k\vert X_{n-1})p_{n-1}(h_0,h_1)^2\,dh_0\,dh_1}\right\}\\
				\times{}&p(X_n=k\vert X_{n-1})p_{n-1}(h_0,h_1)^2\Bigg]\,dh_0\,dh_1.
			\end{aligned}
			\label{eq:two-state_bidirectional_covariance}
		\end{equation}
		\begin{algorithm}[t]
			\caption{Bidirectional two-state chain adaptive Bayesian inference}
			\begin{algorithmic}
				\Require $n=0$, $\theta > 0$, $p_0(h_0,h_1)$ \Comment{$p_0$: prior with support $[0,\infty)^2$.}
				\While{$\text{det} \left( \text{Cov}_n(h_0, h_1) \right) > \theta$}
				\State $n \gets n+1$
				\State $t_n\gets\argmin_{t\ge t_{n-1}} \text{det} \left( \overline{\text{Cov}}_n(h_0, h_1;t) \right) $ \Comment{Calculate $\overline{\text{Cov}}_{n}$ using Eq.~\eqref{eq:two-state_bidirectional_covariance}.}
				\State Draw $\xi_n=(t_n,X_n)$
				\State $p_n(h_0,h_1)\gets \frac{p(\xi_n\vert h_0,h_1)p_{n-1}(h_0,h_1)}{\iint_{\mathbb{R}_{\geq 0}^2} p(\xi_n\vert h_0,h_1)p_{n-1}(h_0,h_1)\,dh_0\,dh_1}$
				\EndWhile
			\end{algorithmic}
			\label{alg:two-state_bidirectional}
		\end{algorithm}
		Note the extra factor of the posterior from the previous sequence of observations $p_{n-1}$ appears to properly weight the probability of transitioning from $X_{n-1}$ to $X_n$. Using the determinant of the expected covariance as computed by Eq.~\eqref{eq:two-state_bidirectional_covariance}, we modify Algorithm~\ref{alg:two-state_unidirectional} to obtain the multi-dimensional adaptive inference algorithm shown in Algorithm~\ref{alg:two-state_bidirectional}. After a sample $\xi_n = (t_n, X_n)$, the resulting covariance is
		\begin{align*}
			\text{Cov}_n(h_i,h_j)=& \iint_{\mathbb{R}_{\geq 0}^2} \Bigg[\left\{h_i-\iint_{\mathbb{R}_{\geq 0}^2} h_i p_{n}(h_0,h_1)\,dh_j \,dh_i \right\} \\
			& \times \left\{h_j-  \iint_{\mathbb{R}_{\geq 0}^2} h_j p_{n}(h_0,h_1) \,dh_i \,dh_j  \right\} p_n(h_0, h_1) \Bigg]\,dh_0\,dh_1.
		\end{align*}
	\subsection{Adaptive Bayesian Inference for Arbitrary Markov Chains}
		While the algorithms proposed above considered chains with two states, we can generalize our approach to arbitrary chains by considering systems with higher dimensional covariance matrices whose determinants we treat as our objective function. 
		Let $X(t)$ be a discrete-state Markov process with $m$ states and $d$ transition rates, denoting $X_n$ as the $n$th state sample, the state $k \in \{ 0, ..., m-1\}$, and the transition rate $h_i$ indexed by $i\in\{1,\dots,d\}$. 
		Note that we could index transition rates as $h_{ij}$ using the ordered pair for the rate of transition from state $X^j \to X^i$, but the single index formulation leads to a more concise form in the terms hereafter. 
		Moreover, we do not always consider Markov chains with complete digraph transition rate conformations that would benefit from ordered pair notation.
	
		To infer the $d$-dimensional vector $\mathbf{h}$ of transition rates, we construct a posterior distribution, for instance, after the $(n-1)$th sample from the sequence $\xi_{1:n-1} = \{(t_1,X_1), ..., (t_{n-1}, X_{n-1})\}$ with a covariance matrix $\Sigma_{n-1} \equiv \text{Cov}_{n-1} \in\mathbb{R}^{d \times d}$ having entries $\left[ \Sigma_{n-1} \right]_{ij} \equiv \text{Cov}_{n-1}(h_i,h_j)$ defining the covariance between the estimates of the transition rates $h_i$ and $h_j$. 
		Our algorithm then proceeds in choosing the next sample time $t_{n}$ that minimizes the determinant of the expected covariance matrix $\text{det}(\overline{\text{Cov}}_n(t))$.
		By marginalizing over possible observable states $X_n \in \{ 0, ..., m-1 \}$, the entries of $\overline{\text{Cov}}_n$, averaged for a particular choice of the next sample time $t_n$, are given by
		\begin{equation}
			\overline{\text{Cov}}_n(h_i,h_j)=\sum_{k=0}^{m-1}\text{Cov}_n(h_i,h_j\vert X_n=k, t_n, \xi_{1:n-1} )\Pr(X_n=k \vert t_n, \xi_{1:n-1}).
			\label{eq:General_Covariance_Marginalization}
		\end{equation}
		As before, we consider the marginalization in Eq.~\eqref{eq:General_Covariance_Marginalization} in the context of transitions from the previous (known) state $X_{n-1}$. 
		This requires introducing the transition probabilities $p\left(X_n=k\vert X_{n-1},\mathbf{h}\right)$ and the posterior $p_{n-1}$ into Eq.~\eqref{eq:General_Covariance_Marginalization}. 
		Formulas for the transition probabilities can be obtained explicitly in a number of cases, but they are not as concise as in the case of two state chains.
		In general, the explicit update rule for the entries of the expected covariance matrix will be
		\begin{equation}
			\begin{aligned}
				{}&\overline{\text{Cov}}_n(h_i,h_j)=\\
				{}&\sum_{k=0}^{m-1}\iint_{\mathbb{R}_{\geq 0}^2} \Bigg[\left\{h_i-\frac{\int_0^\infty h_i\left(\int_0^\infty p\left(X_n=k\vert X_{n-1},\mathbf{h}\right)p_{n-1}(\mathbf{h})^2\,d_i^{d-1}\mathbf{h}\right)\,dh_i}{\int_0^\infty p\left(X_n=k\vert X_{n-1},\mathbf{h}\right)p_{n-1}(\mathbf{h})^2\,d^d\mathbf{h}}\right\}\\
				\times{}&\left\{h_j-\frac{\int_0^\infty h_j\left(\int_0^\infty p\left(X_n=k\vert X_{n-1},\mathbf{h}\right)p_{n-1}(\mathbf{h})^2\,d_j^{d-1}\mathbf{h}\right)\,dh_j}{\int_0^\infty p\left(X_n=k\vert X_{n-1},\mathbf{h}\right)p_{n-1}(\mathbf{h})^2\,d^d\mathbf{h}}\right\}\\
				\times{}&\int_0^\infty p\left(X_n=k\vert X_{n-1},\mathbf{h}\right)p_{n-1}(\mathbf{h})^2\,d^{d-2}_{ij}\mathbf{h}\Bigg]\,dh_i\,dh_j.
			\end{aligned}
			\label{eq:General_Covariance_Update}
		\end{equation}
		In Eq.~\eqref{eq:General_Covariance_Update}, use define the notation $\int\cdot\,d^x_{y_1y_2\dots}\mathbf{z}$ to denote that the integral is $x$-dimensional, and the directions $\{y_1,y_2,\dots\}$ are the directions \emph{not} integrated over.
		For example, $\int\cdot\,d^{d-1}_i\mathbf{h}$ indicates we integrate over all directions in $\mathbf{h}$ except $h_i$.
		Note that for conservation, the number of subindices $y_i$ and the dimension of the integral $x$ must add to the dimension of the space $\mathbf{z}$.
		To use Eq.~\eqref{eq:General_Covariance_Update} to infer a network's transition rates, we substitute this covariance update in place of Eq.~\eqref{eq:two-state_bidirectional_covariance} in Algorithm~\ref{alg:two-state_bidirectional} and modify the normalization step of the posterior appropriately to define Algorithm~\ref{alg:general}.
		In the following, we apply this generalized algorithm to infer transition rates to perform rate inference in some canonical Markov chain models. 
		\begin{algorithm}[t]
			\caption{Adaptive Bayesian inference for general Markov chains}
			\begin{algorithmic}
				\Require $n=0$, $\theta > 0$, $p_0(\mathbf{h})$ \Comment{$p_0$: prior with support $[0,\infty)^d$.}
				\While{$\text{det} \left( \text{Cov}_n(\mathbf{h}) \right) > \theta$}
				\State $n \gets n+1$
				\State $t_n\gets\argmin_{t\ge t_{n-1}} \text{det} \left( \overline{\text{Cov}}_n(\mathbf{h};t) \right) $ \Comment{Calculate $\overline{\text{Cov}}_{n}$ using Eq.~\eqref{eq:General_Covariance_Update}.}
				\State Draw $\xi_n=(t_n,X_n)$
				\State $p_n(\mathbf{h})\gets \frac{p(\xi_n\vert \mathbf{h} )p_{n-1}(\mathbf{h})}{\iint_{\mathbb{R}^d_{\geq 0}} p(\xi_n\vert \mathbf{h})p_{n-1}(\mathbf{h} )\,d \mathbf{h} }$
				\EndWhile
			\end{algorithmic}
			\label{alg:general}
		\end{algorithm}
	\subsection{Numerical Implementation of Adaptive Bayesian Inference Algorithms}
		To numerically implement Algorithms \ref{alg:two-state_unidirectional}-\ref{alg:general}, we must specify a method for finding the optimal sampling times $t_n$ and for updating the posterior density after an observation.
		To solve the optimization problem of finding the sampling time $t_n$ that minimizes the expected covariance $\overline{\text{Cov}}_n$, subject to the constraint $t_n\ge0$, we used MATLAB's \texttt{fminbnd}, which combines golden section search and parabolic interpolation to find optima on a bounded interval. 
		Each successive guess is generated as a convex linear combination of two endpoints of the proposed interval in which the optimum lies \cite{kiefer1953}. 
		We found that standard constrained optimization routines, such as \texttt{fmincon} in MATLAB or \texttt{scipy.optimize.minimize} in Python, provided good results.
		However, we also found that using a bounded optimization routine provided more accurate results for a sufficiently large upper bound on $t_n$.
		To update the posterior density, we first constructed a prior density $p_0$ on a mesh of possible transition rates $\mathbf{h}$, then calculated the posterior update on the same mesh using the exact update rule for the chain configuration (see Eq.~\eqref{eq:transition_probability_matrix_exponential}).
		Because we have access to the exact posterior update rule for any finite-state Markov chain, we were able to avoid the usual computational concerns associated with evaluating posterior densities.
		For further information about our numerical implementations, see \url{https://github.com/nwbarendregt/AdaptMarkovRateInf}.
		
\section{Results: Adaptive Bayesian Inference Applied to Classic Markov Chain Models}
	Having developed our adaptive Bayesian inference algorithm for arbitrary discrete-state Markov chains, we now proceed to study the algorithm's performance on a variety of Markov chain models.
	In all our results, we run the inference algorithm until it has converged, which we define as the first time the  determinant of the posterior covariance (or, in the case of inferring a single transition rate, the posterior variance) drops below a threshold $\theta$.
	To measure algorithm performance, we use two metrics: 1) the number of samples $\xi$ needed for the algorithm to converge, which we define as $N_s$, and 2) the mean-squared error (MSE) of the final posterior output of the algorithm after convergence.
	The MSE associated with a transition rate $h_i$ is defined with respect to the whole posterior, not just the maximum likelihood or the posterior mode:
	\begin{align}
		\text{MSE}_i = \int_0^{\infty} (h_i - h_i^{\text{true}})^2\left[\int_0^\infty p_{N_s}(\mathbf{h})\,d_i^{d-1}\mathbf{h}\right]\, dh_i. \label{mse}
	\end{align}
	In Eq.~\eqref{mse}, $N_s$ is the total number of samples to convergence and $h_i^{\text{true}}$ is the true transition rate, and the integral notation $\int\cdot\,d_i^{d-1}\mathbf{h}$ has the same meaning as in Eq.~\eqref{eq:General_Covariance_Update}.
	Note that we use MSE as a measure of inference error rather than the posterior covariance. Posterior covariance is used to guide sampling time choice and as the convergence criterion since it implies reduction of uncertainty. It is therefore equal across converged algorithms. However, posterior covariance does not necessarily imply how well the algorithm can recover a ground truth model generating observations. In fact, some parameter fitting procedures can lead to biased estimates, in which case uncertainty implied by the parametric posterior is reduced by maximum likelihood estimates stray from the ground truth \cite{smid2020}. This is why we use MSE to evaluate algorithm performance, but this quantity requires knowledge of the true transition rate, which we have access to in our simulation-based studies.
	Alternatively, an experimentalist who wants to study algorithmic performance applied to a given physical system, in which the true transition rates may be unknown, could instead use a performance metric that measures how well the algorithm's output replicates experimental data, such as Bayes factor analysis \cite{kass1995}.
	
	\subsection{Single Transition Rate Inference}
		\label{sec:Simple_Networks_Single_Rate}
			\begin{figure}[t]
			\centering
			\includegraphics[width=\linewidth]{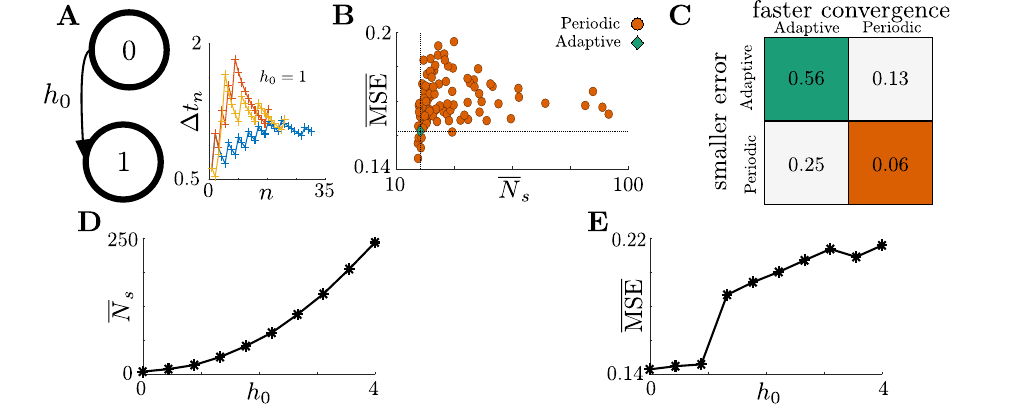}
			\caption{Inferring single transition rates. 
				\textbf{A:} Schematic of two-state Markov process with single transition rate $h_0$. Inset plot shows realizations of the optimal sampling interval computed by the adaptive inference algorithm. \textbf{B:} Scatter plot of algorithms' performance metrics, measured in the number of samples to converge $(\overline{N}_s)$ and average error $(\overline{\text{MSE}})$ for inferring the transition rate $h_0=1$. Green diamond shows performance of the adaptive algorithm, and each orange dot shows the performance of the periodic algorithm for a different fixed sampling period $T$. Periodic algorithm dots that lie in the upper-right quadrant relative to the adaptive algorithm have strictly worse performance than the adaptive algorithm. \textbf{C:} Average performance comparison between adaptive and fixed-period algorithms, taken over $10^3$ values of $h_0$. The periodic algorithm only outperforms the adaptive algorithm $6\%$ of the time. \textbf{D:} Average number of samples to converge $\overline{N}_s$ of adaptive algorithm as a function of fixed transition rate $h_0$. Averages taken over $10^3$ realizations with the same $h_0$. \textbf{E:} $\overline{\text{MSE}}$ of adaptive algorithm as a function of $h_0$, using the same simulations as in \textbf{D}.}
				\label{fig:1}
		\end{figure}
		We compared the performance of adaptive inference for the two-state, single transition rate chain schematized in Fig.~\ref{fig:1}\textbf{A}, defined by Algorithm~\ref{alg:two-state_unidirectional}, to that of an algorithm using a predetermined, fixed sampling period $T$, which for a general Markov chain is given by Algorithm~\ref{alg:fixed_period}.
		\begin{algorithm}[b]
			\caption{Fixed-Period Inference Algorithm}
			\begin{algorithmic}
				\Require $n=0$, $\theta > 0$, $p_0(\mathbf{h})$, $T>0$ \Comment{$p_0$: prior with support $[0,\infty)^d$.}
				\While{$\text{det} \left( \text{Cov}_n(\mathbf{h}) \right) > \theta$}
				\State $n \gets n+1$
				\State $t_n\gets t_{n-1}+T$ \Comment{Update sample time using fixed-period input $T$.}
				\State Draw $\xi_n=(t_n,X_n)$
				\State $p_n(\mathbf{h})\gets \frac{p(\xi_n\vert \mathbf{h})p_{n-1}(\mathbf{h})}{\iint_{\mathbb{R}_{\geq 0}^d} p(\xi_n\vert \mathbf{h})p_{n-1}(\mathbf{h})\,d\mathbf{h}}$
				\EndWhile
			\end{algorithmic}
			\label{alg:fixed_period}
		\end{algorithm}
		Randomly selecting parameters from a gamma-distributed prior with hyperparameters $(\alpha,\beta)=(2,1)$, we determined the average number of samples for both algorithms to converge and the average mean-squared error (MSE) of both algorithms once they have converged.
		Note that periodic algorithm's performance will depend on the choice of the fixed sampling period $T$; for the adaptive inference algorithm, which does not take $T$ as an input, performance will be independent of $T$. 
		Comparing the performance of both algorithms across a range of values of $T$, we found that the adaptive algorithm tended to strongly outperform the periodic algorithm in convergence time and inference error.
		This performance advantage can be clearly observed both when inferring a single ground truth transition rate parameter (Fig.~\ref{fig:1}\textbf{B}) as well as across many realizations of transition rates (Fig.~\ref{fig:1}\textbf{C}).
		Additionally, the adaptive algorithm inherently identifies the best choice of the sampling time for each sample based on the posterior over transition rates inferred so far, integrating the process of parameter inference with sampling method as an online experimental design.
		We also measured the number of samples required and the MSE of the adaptive algorithm for fixed $h_0$ to investigate any systematic biases in adaptive inference. Inferring larger transition rates requires more samples (Fig.~\ref{fig:1}\textbf{D}) due to (a) the sensitivity of the posterior to state observations at short observation times and (b) low prior likelihood due to their place in the tail of the gamma distribution. Inference error, defined as the MSE between the posterior and true transition rate, was uniformly low across all values of $h_0$ (Fig.~\ref{fig:1}\textbf{E}). Note that termination at low posterior variance does not ensure low MSE, since observations could guide the mean estimate away from the true value, and Bayesian inference of parameters with insufficient sample sizes or priors can be strongly biased \cite{smid2020}.

	\subsection{Higher-dimensional Inference on Two-State Chains}
\begin{figure}[t]
			\centering
			\includegraphics[width=\linewidth]{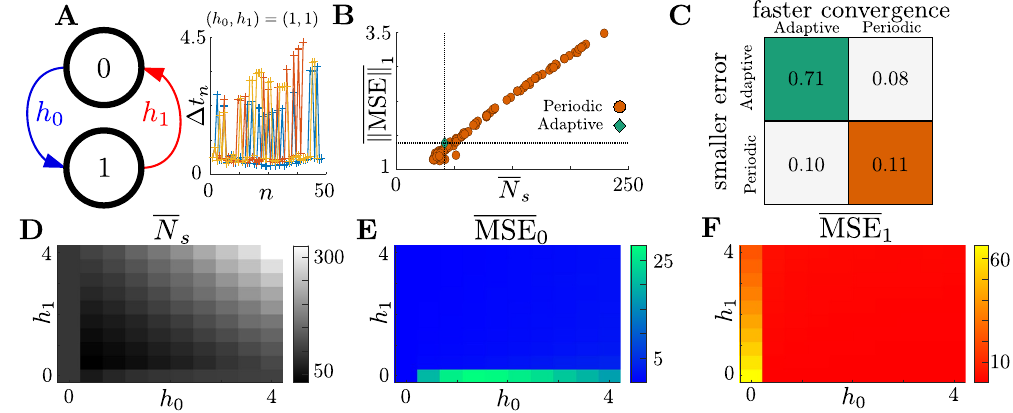}
			\caption{Inferring multiple transition rates. 
				\textbf{A:} Schematic of two-state Markov process with two transition rates $h_0$, $h_1$ (originating from state 0 and 1, respectively). Inset plot shows realizations of the optimal sampling interval computed by the adaptive inference algorithm. Note the distinct dichotomous nature of the set of time choices. \textbf{B:} Scatter plot of algorithms' performance metrics, measured in the number of samples to converge $(\overline{N}_s)$ and average norm of the error $(\overline{\Vert\text{MSE}\Vert}_1)$ for inferring the transition rate pair $(h_0,h_1)=(1,1)$. Green diamond shows performance of the adaptive algorithm, and each orange dot shows the performance of the periodic algorithm for a different fixed sampling period $T$. Periodic algorithms that lies in the upper-right quadrant relative to the adaptive algorithm have strictly worse performance than the adaptive algorithm. \textbf{C:} Average performance comparison between adaptive and fixed-period algorithms, taken over $10^3$ pairs of $(h_0,h_1)$. The periodic algorithm only outperforms the adaptive algorithm $11\%$ of the time. \textbf{D:} Mean number of samples $\overline{N}_s$ required for adaptive algorithm to reach the covariance determinant threshold as a function of fixed true transition rates $h_0$, $h_1$. Averages taken over $10^3$ realizations with the same pair of transition rates. \textbf{E:} $\overline{\text{MSE}}$ for $h_0$ inference of adaptive algorithm as a function of $h_0$, $h_1$, taken using the same simulations as in \textbf{D}. \textbf{F:} Same as \textbf{E}, but for $\overline{\text{MSE}}$ for $h_1$ inference.}
			\label{fig:2}
		\end{figure}

		To assess the adaptive algorithm's performance performance on a slightly higher-dimensional problem, we considered a two-state Markov chain with two transition rates (schematized in Fig,~\ref{fig:2}\textbf{A}) and compared Algorithm~\ref{alg:two-state_bidirectional} to a fixed-period sampling algorithm, identifying the best period for a given prior.
		To mirror the gamma prior used previously, we chose a bivariate gamma prior with joint distribution function \cite{nadarajah2006}
		\begin{equation}
			\begin{aligned}
				p_0(h_0,h_1)={}&C\Gamma(b)(h_0h_1)^{c-1}\left(\frac{h_0}{\mu_1}+\frac{h_1}{\mu_2}\right)^{\frac{a-1}{2}-c}\exp\left\{-\frac{1}{2}\left(\frac{h_0}{\mu_1}+\frac{h_1}{\mu_2}\right)\right\}\\
				\times{}&W_{c-b+\frac{1-a}{2},c-\frac{a}{2}}\left(\frac{h_0}{\mu_1}+\frac{h_1}{\mu_2}\right),
			\end{aligned}
			\label{eq:bivariate_gamma_pdf}
		\end{equation}
		where $C$ is given by
		\begin{align*}
			\frac{1}{C}=\left(\mu_1\mu_2\right)^c\Gamma(c)\Gamma(a)\Gamma(b),
		\end{align*}
		$c=a+b$, and $W$ is the Whittaker function given by
		\begin{align*}
			W_{\lambda,\mu}(a)=\frac{a^{\mu+\frac{1}{2}}e^{-\frac{a}{2}}}{\Gamma\left(\mu-\lambda+\frac{1}{2}\right)}\int_0^\infty t^{\mu-\lambda-\frac{1}{2}}(1+t)^{\mu+\lambda-\frac{1}{2}}e^{-at}\,dt.
		\end{align*}
		We will take $\mu_1=\mu_2=2$ and $a=b=1$ throughout when using the bivariate gamma prior given by Eq.~\eqref{eq:bivariate_gamma_pdf}.
		Sampling different pairs of transition rates $(h_0,h_1)$ from this prior, we compared the convergence time and inference error of both algorithms using the same approach we implemented for the single-transition problem (Fig.~\ref{fig:2}\textbf{B},\textbf{C}).
		For this higher-dimensional inference problem, the adaptive algorithm still showed better performance across a range of different transition rate pairs.
		Because an experimentalist cannot know the best sampling period a priori, these results suggest the adaptive algorithm is generally faster and more accurate than the na\"ive approach.
			
		Does increasing the dimension of the inference problem introduce any new biases in adaptive inference?
		We measured average convergence time (Fig.~\ref{fig:2}\textbf{D}) and average inference errors (Fig.~\ref{fig:2}\textbf{E},\textbf{F}) of the adaptive inference algorithm for fixed transition rates using the same bivariate gamma prior.
		Similar to the single-transition network in Section~3\ref{sec:Simple_Networks_Single_Rate}, inferring larger transition rates takes more samples, as these rates are in the tails of the bivariate gamma prior and observables (state sequences) are less sensitive to subtle changes in parameters (transition rates).
		Additionally, the average MSE associated with each rate is consistently low across most of parameter space.
			
		The notable exception to this low-error behavior is when one of the transition rates ($h_i$) is identically zero in which case the estimate of the other transition rate ($h_j$) is poor.
		To understand this behavior, consider a network where $h_0 = 0$ and $h_1 \ge 0$.
		If the initial state is $X_0 = 0$, then no transitions will occur.
		Obtaining the same $X_n=0$ measurements implies either: 1) the rate $h_0$ is small, or 2) the rate $h_1$ is large.
		The posterior of the adaptive algorithm converges to account for both of these possibilities, which results in small errors in $h_0$ inference and potentially large errors in $h_1$ inference.
		Heuristically, we can explain this behavior by noting that repeated $X_n=0$ measurements means that, in the large-sample limit and for fixed intersample interval $\delta t$, the posterior scales as powers of the first term in Eq.~\eqref{eq:two-state_bidirectional_transition_probs}:
		\begin{equation*}
			p_n(h_0,h_1)\propto\left[\frac{h_1}{h_0+h_1}+\frac{h_0}{h_0+h_1}e^{-(h_0+h_1)\delta t}\right]^n.
		\end{equation*}
		Fixing $h_1\ge0$, this posterior is maximized at $h_0=0$, so as $n$ increases, $p_n$ will asymptotically converge to a delta distribution along the $h_0$ dimension at $h_0=0$.
		Simultaneously, for fixed $h_0\to0$, $p_n$ appears as a flat distribution along the $h_1$ dimension, implying the posterior contains no information about $h_1$.
		These results demonstrate that achieving accurate inference requires utilizing a prior with dimension equal to the dimension of the problem.
		\subsubsection{Effect of Convergence Tolerance on Adaptive Inference}
			\label{sec:Tolerance_Refinement}
\begin{figure}[t]
				\centering
				\includegraphics[width=\linewidth]{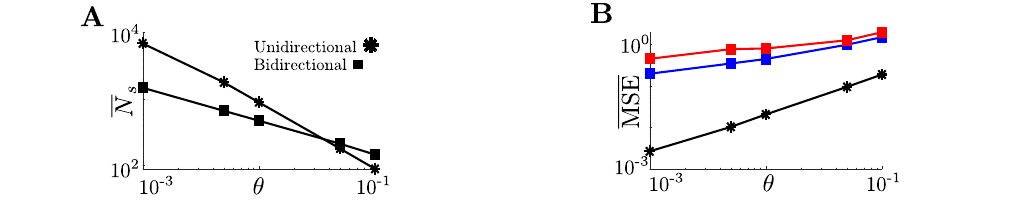}
				\caption{Performance of adaptive algorithm for varied convergence tolerances.
					 \textbf{A:} Average number of samples $\overline{N}_s$ required for the adaptive algorithm to converge for different tolerances $\theta$ applied to both two-state networks in Section~\ref{sec:Simple_Networks} (legend). Averages taken over $10^3$ transition rates drawn from a $\Gamma(2,1)$ prior for the unidirectional network and $10^3$ pairs of rates from the bivariate gamma prior, Eq.~\eqref{eq:bivariate_gamma_pdf}, for the bidirectional network. \textbf{B:} Mean squared error $\overline{\text{MSE}}$ of adaptive algorithm with varied $\theta$ applied to the same networks and sampled transition rates as in \textbf{A}.}
				\label{fig:3}
			\end{figure}

			So far we have investigated the adaptive algorithm's convergence speed and inference error for a fixed convergence tolerance $\theta$. However, the choice of $\theta$ may impact the algorithm's performance.
			Using the simple two-state Markov processes discussed above, we measured the convergence time (Fig.~\ref{fig:3}\textbf{A}) and inference error (Fig.~\ref{fig:3}\textbf{B}) of the adaptive inference algorithm as $\theta$ is varied.
			Changing $\theta$ leads to a trade-off between the error in the estimate and the number of samples required for convergence: lower (tighter) tolerances require more samples to converge, but yield low-error estimates.
			Comparing the algorithm's performance between unidirectional versus bidirectional Markov chains, we found that inferring multiple transition rates requires fewer samples than inferring a single transition for nearly all values of $\theta$.
			This trend is likely a result of the increase in the inference problem's dimension: convergence in the multi-dimensional inference algorithm is measured using covariance as opposed to variance, and minimizing the determinant of the covariance can be achieved both by minimizing the individual variances and by maximizing the correlations between the two rate estimations.
	\subsection{Multi-dimensional Inference on Complex Chains}
		\subsubsection{Inferring Birth and Death Rates in an M/M/1 Queuing Process}
\begin{figure}[t]
				\centering
				\includegraphics[width=\linewidth]{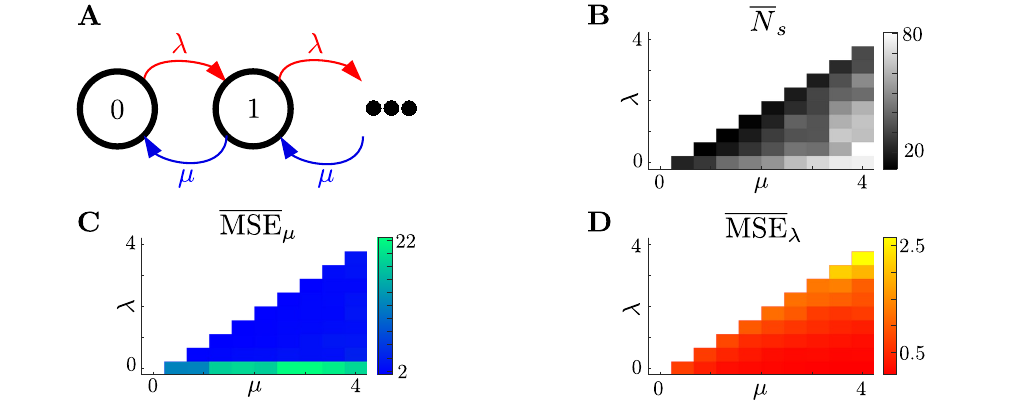}
				\caption{Inferring transition rates in a basic queueing process.
					\textbf{A}: Schematic of a M/M/1 queue process with ``birth rates'' $\lambda$ and ''death rates'' $\mu$.
					\textbf{B}: Average number of samples $\overline{N}_s$ required for the adaptive algorithm to infer fixed transition rates $\mu$, $\lambda$, with restriction $\lambda < \mu$.
					Average taken over $10^2$ realizations with the same pair of transition rates.
					\textbf{C}: Mean squared error $\overline{\text{MSE}}$ estimating $\mu$, inferred using the adaptive algorithm for fixed transition rates, taken over the same samples as in \textbf{B}.
					\textbf{D}: Same as \textbf{C}, but for $\overline{\text{MSE}}$ estimating $\lambda$.}
				\label{fig:4}
			\end{figure}

			We start by considering an M/M/1 queue (birth-death process) Markov chain $X(t)\in\{0,1,2,\dots\}$ ($m \to \infty$), schematized in Fig.~\ref{fig:4}\textbf{A}, with birth rates $\lambda\ge0$ and death rates $\mu>0$, with the restriction $\mu > \lambda$ for boundedness, so that
			\begin{gather*}
				\Pr(X(t+\delta t)=i+1\vert X(t)=i)=\lambda\delta t+o\left(\delta t\right), \quad \forall i\ge0,\\
				\Pr(X(t+\delta t)=i-1\vert X(t)=i)=\mu\delta t+o\left(\delta t\right), \quad \forall i\ge1,\\
				\Pr(X(t+\delta t)=i\vert X(t)=i)=1-(\lambda+\mu)\delta t+o\left(\delta t\right), \quad \forall i\ge0.
			\end{gather*}
			One can show (see \cite{gross1998} for details) that the transition probabilities $p(X(t_n)=j\vert X(t_{n-1})=i,\lambda,\mu)$ for this M/M/1 queue are given by the formula involving the time interval $\Delta t = t_n - t_{n-1}$ and states $i$ and $j$:
			\begin{equation}
				\label{eq:MM1_transition_probability}
				\begin{aligned}
					{}&p(X(t_n)=j\vert X(t_{n-1})=i,\lambda,\mu)=e^{-(\lambda+\mu)\Delta t}\\
					\times{}&\left\{\rho^{\frac{j-i}{2}}I_{j-i}(a\Delta t)+\rho^{\frac{j-i-1}{2}}I_{j+i+1}(a\Delta t)+(1-\rho)\rho^j\sum_{k=j+i+2}^\infty \rho^{-\frac{k}{2}}I_k(a\Delta t)\right\},
				\end{aligned}
			\end{equation}
			where $\rho=\frac{\lambda}{\mu} < 1$ by the condition placed on the birth/death rates, $I_k$ is the modified Bessel function of the first kind, and $a=2\sqrt{\lambda\mu}$.
			Substituting Eq.~\eqref{eq:MM1_transition_probability} into Eq.~\eqref{eq:General_Covariance_Update}, we can proceed with inferring the transition rates $\lambda$ and $\mu$ using Algorithm~\ref{alg:general}.
			As mentioned, restricting possible transition rates so that $\mu>\lambda$ guarantees a well-defined and finite stationary distribution for the Markov chain.
			We accomplish this restriction by drawing the birth $\lambda$ and death $\mu$ rates from a truncated version of the bivariate gamma prior from Eq.~\eqref{eq:bivariate_gamma_pdf} which ensures death rates are always larger.

			To determine how adaptive inference fares in specifying the birth and death rates of this countably infinite Markov chain, we again quantify estimation error and convergence time.
			Across all transition rates considered, the adaptive algorithm quickly converges to an accurate estimate of the transition rates (Fig.~\ref{fig:4}\textbf{B}).
			As in the simple networks discussed above, the algorithm converges slower when the transition rates are in the tails of the prior.
			Additionally, as in the two-state network (Fig.~\ref{fig:2}\textbf{E}), the algorithm has poor accuracy for inferring $\mu$ when $\lambda=0$ (Fig.~\ref{fig:4}{\bf C}).
			For this pure death process, the chain will always eventually converge to the absorbing state $X=0$ for all $\mu>0$, yielding little information about the magnitude of $\mu$ itself.
			However, unlike the two-state network (Fig.~\ref{fig:2}\textbf{F}), the algorithm had very low error in inferring $\lambda$ for all values considered (Fig.~\ref{fig:4}{\bf D}). The adaptive algorithm is effective in inferring the parameters of this birth-death process, particularly when the two parameters have similar value but the death rate is still larger than the birth rate.
		\subsubsection{Inferring Transition Rates in a Ring Network}
\begin{figure}[t]
				\centering
				\includegraphics[width=\linewidth]{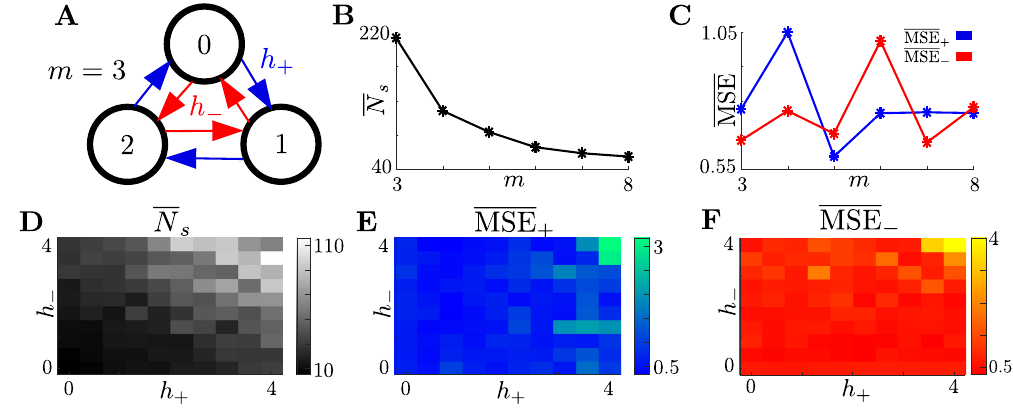}
				\caption{Inferring transition rates in a symmetric ring network. 
					\textbf{A}: Schematic of a ring network of size $m=3$ with clockwise transition rates $h_+$ and counterclockwise transition rates $h_-$.
					\textbf{B}: Average number of samples $\overline{N}_s$ required for the adaptive algorithm to infer transition rates given a ring of size $m$.
					Averages taken over $10^2$ sampled pairs of transition rates generated from the bivariate gamma prior, Eq.~\eqref{eq:bivariate_gamma_pdf}.
					\textbf{C}: Mean squared error $\overline{\text{MSE}}$ in estimating the transition rates of the ring chain using the adaptive algorithm for varied network size $m$ and the same network realizations from \textbf{B}.
					\textbf{D}: Mean number of samples $\overline{N}_s$ required for the adaptive algorithm's estimate of the transition rates to converge in a network of size $m=8$ for a fixed pair $(h_-, h_+)$ of true counterclockwise and clockwise transition rates.
					Averages at each parameter set value taken over $10^2$ trials.
					\textbf{E}: Mean squared error $\overline{\text{MSE}}$ in the estimate of $h_+$ as the two transition rates are varied.
					Averages taken using the same trials as in \textbf{D}.
					\textbf{F}: Same as \textbf{E}, but for $\overline{\text{MSE}}$ estimating $h_-$.}
				\label{fig:5}
			\end{figure}

			Every chain we have considered so far has shared a key feature: we can introduce an absorbing state in the chain by setting one of the transition rates to zero. Cutting a single link of the chain causes one state to have no link out of it. 
			This can lead to high errors in inference for both the two-state (Fig.~\ref{fig:2}\textbf{E},\textbf{F}) and M/M/1 queue (Fig.~\ref{fig:4}\textbf{C}) networks.
			To move away from these cases, we consider a ring chain with identical clockwise transition rates $h_+$ and counterclockwise transition rates $h_-$ (Fig.~\ref{fig:5}\textbf{A}).
			This chain, which is equivalent to a periodic random walk, possesses absorbing states only if $h_+$ and $h_-$ are identically zero, so we can further test if adaptive inference error increases due to the reduction in a problem's dimension.
			For such a ring network with $m$ states, the transition probabilities from state $j$ to $i$ given a time interval and the clockwise and counterclockwise transition rates, $p(X(t_n)=j\vert X(t_{n-1})=i,h_+,h_-)$, are given by the matrix exponential \cite{taylor1984}
			\begin{equation}
				p(X(t_n)=j\vert X(t_{n-1})=i,h_+,h_-)=\left[e^{\mathbf{A}(h_+,h_-)\Delta t}\right]_{ij},
				\label{eq:transition_probability_matrix_exponential}
			\end{equation}
			where $\Delta t=t_n-t_{n-1}$ and $\mathbf{A}\in\mathbb{R}^{m \times m}$ is the infinitesimal generator matrix for the network.
			In the case of an $m$-state ring network, $\mathbf{A}(h_+,h_-)$ is given by
			\begin{align*}
				\mathbf{A}=\begin{pmatrix}
					-(h_++h_-) & h_+ & 0 & \dots & 0 & h_-\\
					h_- & -(h_++h_-) & h_+ & 0 & \dots & 0\\
					0 & \ddots & \ddots & \ddots & & \\
					\vdots & & & & & \vdots \\
					& & & & & 0\\
					\vdots & \dots & 0 & h_- & -(h_++h_-) & h_+\\
					h_+ & 0 & \dots & 0 & h_- & -(h_++h_-) 
				\end{pmatrix}.
			\end{align*}

			We were first interested in how the size of the Markov chain affected the error in inference and the time for the algorithm to converge.
			Recall, inference of the transition rate parameters converged faster in the M/M/1 queue chain than in the simple two-state chain, which we attributed to an increase in the number of chain states (i.e., increase in the possible measurements), allowing for a more refined sampling of the stochastic dynamics of the chain with each observation.
			We looked to see if this trend extended to the context of a chain with ring topology, computing the convergence time (Fig.~\ref{fig:5}\textbf{B}) and error (Fig.~\ref{fig:5}\textbf{C}) in transition rate estimation as a function of chain size.
			The results suggest that increasing the chain size does in fact speed up convergence of the adaptive algorithm. Moreover, this speed-up does not generate any additional error in rate inference, as the adaptive algorithm displays uniformly low error for all chain sizes considered.  

			We also looked further at how strong asymmetries in the true transition rate parameters impacted the performance of the adaptive algorithm on the rate inference problem on the ring. To do so, we fixed the ring size $m=8$ and measured convergence time (Fig.~\ref{fig:5}\textbf{D}) and inference errors of each transition rate (Fig.~\ref{fig:5}\textbf{E},\textbf{F}) for different fixed pairs of $h_+$ and $h_-$.
			As we observed in all previous chains, the adaptive algorithm takes longer to converge when the true transition rates are large and fall in the tails of the bivariate gamma prior (Eq.~\eqref{eq:bivariate_gamma_pdf}).
			Unlike the previous examples we studied, the average inference error for both transition rates is uniformly low across the range of values considered.
			These findings provide further credence to our speculation that the appearance of absorbing states in Markov chains can drastically increase the error in rate parameter inference. The only way for the ring chain to become absorbing would be for both transition rates to be identically zero, causing the system to be frozen in the initial state from the start. Otherwise, the observations of the Markov chain are guaranteed to span the entire state space and continually provide new information about both transition rates to the adaptive algorithm.
			Additionally, because the only absorbing network occurs when $h_+=h_-=0$, the algorithm quickly infers this configuration by obtaining repeated measurements of the same state, so even these parameters can be rapidly inferred to high precision.
		\subsubsection{Inferring Network Structure}
\begin{figure}[t]
				\centering
				\includegraphics[width=\linewidth]{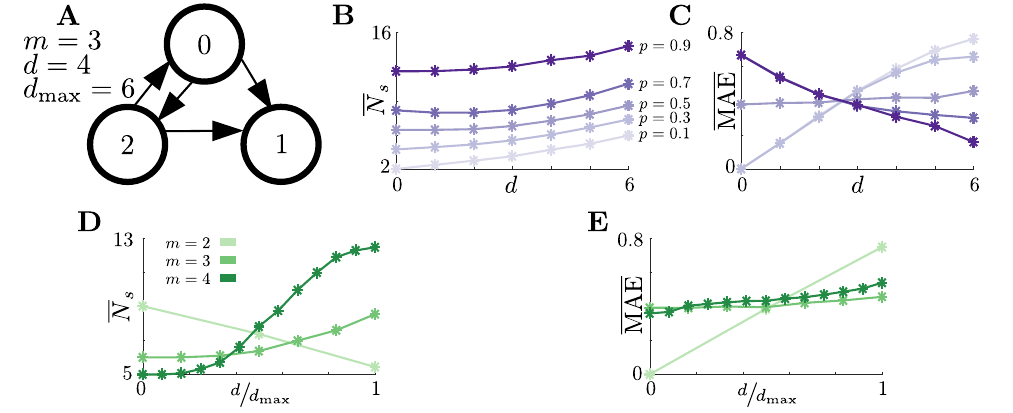}
				\caption{Inferring structure of Markov chains with rates drawn from binary sets.
					\textbf{A}: Schematic of a sample binary chain of size $m=3$ with $d=4$ nonzero transition rates. All transition rates are each independently chosen from the set $\{0,1\}$ with Bernoulli parameter $p$ (see text for details).
					\textbf{B}: Average number of samples $\overline{N}_s$ needed for the adaptive algorithm to infer the transition rates to the required degree of accuracy, defined by Eq.~(\ref{MAE}), as a function of the number of nonzero transition rates $d$ with fixed network size $m=3$.
					Averages taken over $10^3$ sampled network structures with independent Bernoulli parameter $p$; several values of $p$ are superimposed (labeled).
					\textbf{C}: Average inference error, measured using mean absolute error ($\overline{\text{MAE}}$, $L_1$-error), for varied $d$.
					Averages computed using the same trials as in \textbf{B}.
					\textbf{D}: Average number of samples $\overline{N}_s$ required for the adaptive algorithm to converge to a set accuracy for different chain sizes (legend).
					Averages taken over $10^3$ sampled network structures with fixed Bernoulli prior $p=0.5$.
					\textbf{E}: Mean absolute error $\overline{\text{MAE}}$ to which the adaptive algorithm converges as the density of links in the chain is increased for several different network sizes.
					Averages computed using the same data from \textbf{D}.}
				\label{fig:6}
			\end{figure}

			As a final test of the adaptive algorithm, we considered the problem of inferring the structure of a Markov chain with a strongly restrictive prior on the rates, requiring that they are either 0 or 1. Doing so isolates the problem of identifying the presence or absence of a link in the Markov chain without the further problem of inferring the amplitude of the rate.
			Thus, each transition rate is a binary variable drawn independently from the set $\{0,1\}$ with Bernoulli parameter $p$ (Fig.~\ref{fig:6}\textbf{A}).
			In this way, we reduce the set of possible chain link conformations by only allowing for one possible non-zero value for all transition rates.
			The transition probabilities $p(X(t_n)=j\vert X(t_{n-1})=i,\mathbf{h})$ for these networks are given by the same matrix exponential as in the case of ring chains, described in Eq.~\eqref{eq:transition_probability_matrix_exponential}, where the infinitesimal generator matrix $\mathbf{A}$ is changed to reflect the specific chain's structure. 
			To measure our algorithm's performance on this class of chains, we compute the average number of samples required to converge and, due to the binary prior over the transition rates, measure inference using mean absolute error (MAE), given by a normalized $L_1$-error
			\begin{equation}
				\text{MAE}=\frac{\sum_{k=1}^{d_\text{max}}\vert h_k-\hat{h}_k\vert}{d_\text{max}}, \label{MAE}
			\end{equation}
			where $d_\text{max}=m(m-1)$ is the maximum number of possible nonzero transition rates for a chain of size $m$, $h_k$ is the true value of the $k$-th transition rate, and $\hat{h}_k$ is the maximum a posteriori estimate of the $k$-th transition rate.
			To account for the fact that the initial determinant of the covariance matrix changes as $d_\text{max}$ increases, we modified the convergence threshold to depend on this initial covariance determinant.
			For example, if the initial determinant for a simulation was $D$, we ran our adaptive algorithm until the covariance determinant was less than $\theta D$.
			We took $\theta=10^{-2}$ for all simulations on these binarized networks.

			How does adaptive inference handle this problem?
			For a fixed chain size, inferring the structure is faster when the chain is more disconnected (Fig.~\ref{fig:6}\textbf{B}). More connected networks provide a higher diversity of possible measurements, increasing the possible number of network configurations that may generate those measurements.
			Note that because these networks have binary transition rates, our algorithm does not have to infer the magnitude of a non-zero transition rate and therefore avoids the errors shown in Fig.~\ref{fig:2}\textbf{E},\textbf{F}. 
			Additionally, inference error heavily depends on both the true connectivity of the network and the prior likelihood over each transition rate (Fig.~\ref{fig:6}\textbf{C}).
			However, error is lowest when the true connectivity is aligned with the prior (i.e., when both $\frac{d}{d_\text{max}}<0.5$ and $p<0.5$ or $\frac{d}{d_\text{max}}>0.5$ and $p>0.5$), and increases when the two are mismatched (e.g., when $p$ is closer to 1 but $d$ is closer to 0). 
			For a fixed prior, rate estimate performance displays similar behaviors across different chain sizes if there are more than two states: the algorithm converges faster when chains are more disconnected (Fig.~\ref{fig:6}\textbf{D}), and error is lower when the chain structure and the prior agree (Fig.~\ref{fig:6}\textbf{E}).
			These performance trends are consistent with our findings from applying adaptive inference to other Markov chain models: transition rates in the tails of the prior or that are highly asymmetric take more measurements to infer, but generally have low inference error.
			When some transition rates are set to zero, creating absorbing states in the chain, inferring the existence of non-zero transition rates is less difficult than inferring the magnitude of those rates.
			However, across a range of chain structures and transition rate magnitudes, adaptive inference is able to rapidly and accurately infer state transition dynamics.
\section{Conclusions}
In this work, we developed a simple algorithm to infer transition rates of arbitrary discrete-state Markov processes that determines optimal sampling times to minimize a posterior covariance.
Starting with small chains made up of two states, we found that using a previously-developed adaptive algorithm by \cite{webb2021} had lower error than a na\"ive algorithm that samples with a fixed period.
Because of the simplicity of our approach, we showed how to extend the adaptive algorithm to infer generic structures of transition rates, where sample times are chosen to minimize the determinant of the posterior covariance matrix.
Applying this extension to more complex Markov chains, we found that the adaptive algorithm rapidly converged to accurately estimate rate parameters when the true chain structure was more likely according to the prior.
When the chain link conformation was less likely according to the prior, the adaptive algorithm still converged fairly quickly, but inference error was higher for one or more of the transition rates. 

Our work builds on previous development of sequential Bayesian methods for experimental design, harnessing stepwise posterior updates to guide design parameter choices.
		In addition to the work in traditional experimental design, our framework is also closely related to standard approaches used in active learning \cite{mackay1992}, and leverages advances in classic Bayesian optimization problems \cite{shahriari2015}.
		However, when inferring complex models, these fields are often faced with likelihood functions that quickly become intractable and make posterior updates challenging, much like the problems facing the experimental design studies we have previously mentioned.
		Here, we demonstrate that for the large class of models that can be described using Markov chains, adaptive Bayesian design can be implemented to great effect, has analytically-tractable observational likelihoods, and does not require specialized, computational techniques to perform posterior updates. This moves considerably beyond a recent study of adaptive sampling for Markov chains~\cite{michel2020}, which would update the sampling time after several samples rather than each time, and only ensured asymptotic equivalence to optimal fixed time designs.

While we only considered Markovian networks with constant transition rates, our posterior-covariance-minimization approach can theoretically be extended to transition rates of arbitrary functional forms.
These extensions could prove useful for inferring transition rates in stochastic systems with variable rates, as found in chemical kinetic systems modeled by M/G/1 queueing processes \cite{anderson2015} and stochastic implementations of Hodgkin-Huxley neuronal dynamics with voltage-dependent transition rates \cite{pu2020,pu2021}.
Our adaptive inference algorithm can also be adapted to more complex chain structures, such as those used in age-structured epidemiological models \cite{zhang2021} and models for chaperone-assisted protein folding \cite{ilker2021}.
The only hard constraints to our approach are that the possible transition functions be fully specified and the chain itself be Markovian.
However, our algorithm cannot escape the curse of dimensionality for more complex chains. For a Markov chain with $d$ distinct transition rates (or equivalently, $d$ parameters that specify the transition rate functions) and a numerical discretization that allows each transition rate to take on $n$ possible values, the size of the posterior grows as $n^d$.
Future work utilizing our algorithm for more complex transition rate inference problems would necessitate efficient matrix methods or posterior approximation techniques. 
\vskip6pt

\enlargethispage{20pt}

\dataccess{For the MATLAB code used to generate all results and figures, see\\  \url{https://github.com/nwbarendregt/AdaptMarkovRateInf}.}

\competing{The authors declare that they have no competing interests.}

\funding{This work was supported by CRCNS/NIH R01-MH-115557, NIH R01-EB029847-01, and NSF-DMS-1853630.}

\bibliographystyle{RS}
\bibliography{Rate_Inference_References}
\end{document}